# Universal theory of multipartite cavity quantum electrodynamics

Moslem Alidoosty Shahraki and Sina Ataollah Khorasani

*School of Electrical Engineering, Sharif University of Technology*

*P. O. Box 11365-9363, Tehran, Iran; E-mail: khorasani@sina.sharif.edu*

**Abstract:**

Cavity quantum electrodynamics of multipartite systems is studied in depth, which consist of an arbitrary number of emitters in interaction with an arbitrary number of cavity modes. The governing model is obtained by taking the full field-dipole and dipole-dipole interactions into account, and is solved in the Schrödinger picture without assumption of any further approximation. An extensive code is developed which is able to accurately solve the system and track its evolution with high precision in time, while maintaining sufficient degrees of arbitrariness in setting up the initial conditions and interacting partitions. Using this code, we have been able to numerically evaluate various parameters such as probabilities, expectation values (of field and atomic operators), as well as the concurrence as the most rigorously defined measure of entanglement of quantum systems. We present and discuss several examples including a seven-partition system consisting of six quantum dots interacting with one cavity mode. We observe for the first time that the behavior of quantum systems under ultrastrong coupling is significantly different than the weakly and strongly coupled systems, marked by onset of a chaos and abrupt phase changes. We also discuss how to implement spin into the theoretical picture and thus successfully simulate a recently reported spin-entanglement experiment.

## I. INTRODUCTION

Without doubt, Cavity Quantum Electrodynamics (CQED) is one of the frontiers of modern science, where its applications are rapidly entering the realm of engineering fields. With the advent of quantum computing, CQED remains at the cutting edge of the technology in this area, as the most successful and only commercialized platform. This is while the increasing number of quantum sub-systems, or the so-called partitions, demand for more complicated analytical and simulation tools capable of dealing with multipartite systems without losing accuracy. As the number of partitions increase, the difficulty in treatment of multipartite systems unfolds in two aspects: (a) how to maintain accuracy while increasing the system dimensionality, and (b) how to write down equations using proper notations and mathematical expression without causing confusion and/or ambiguity. This is while we have to add the computational complexity of solving algorithm unaddressed.

The first successful theoretical understanding of CQED was made by Jaynes and Cummings in 1963 [1,2] and independently by Paul [3], as explicit solutions to the so-called Jaynes-Cummings-Paul Model (JCPM), for two-level emitters interacting resonantly or non-resonantly with one radiation mode. Soon after, JCPM proved its usefulness in description of spontaneous emission [4], and collapse and revival of wavefunctions [5,6]. Further development in experimental techniques and preparation of Rydberg atoms having very large transition dipole moments allowed various interaction regimes to be studied with regard to the magnitude of coupling constant or Rabi frequency [7,8]. According to the strength of this constant, a CQED system may fall in three regimes: weakly, strongly, or ultrastrongly coupled. If the Rabi frequency is less than the decay rate of excited states in the cavity then the interaction is usually weak, which happens to address most of the occurring CQED systems. Weakly coupled systems are responsible for a number of well-known phenomena such as enhanced or suppressed spontaneous radiation, which have found applications in modern light emitting devices such as semiconductor lasers [9,10].

Larger coupling constants exceeding one-fourth of the difference of atom and field decay rates [11], enables the strong coupling regime. Under strong coupling, the eignstates would be no longer degenerate and recombine to form two separate state pairs [12,13], having an energy difference given by the Rabi frequency. If the emitting system is a quantum dot or well, then these new quantum mixed states obtained from photon-exciton interaction are sometimes viewed as new quasi-particles referred to as exiton-polaritons. Strongly coupled systems are builing blocks of solid-state quantum information processing [14], and novel quantum phenomena including anti-bunching in single-photon emitters [15], quantum encryption [16], quantum repeaters [17,18], and quantum computation [19].

Strong coupling in a cavity may be reached by increasing confinement times and thus quality factors, while decreasing the effective mode volume [20]. At optical frequencies, semiconductor cavities [21], micro-disks [22] and photonic crystal nano-defects [23-27] have demonstrated successful operation of ultra-low threshold and polariton lasers, through combining excellent confinement and tight mode volumes. Another aspect of strongly-coupled CQED systems is control of detuning frequency. High-fidelity single photon sources [28-31], high bandwidth low-threshold lasers [32-34], and single quantum-dot devices such as mirrors [35] and phase shifters [36] are highly dependent on the possibility of control on detuning. Among these, various methods such as cryogenic lattice temperature control [21,37,38], condensation at ultralow temperatures [39,40] and electrical control [41] may be mentioned.

The next interaction regime is the ultrastrong coupling, where the Rabi frequency is typically comparable or even larger than the decay rate [42]. This typically results in improved excited and ground state properties such as non-adiabatic CQED phenomena [42]. Most ultrastrongly coupled systems depend on the range of the radiation spectrum occur in either of the two solid state systems. At the optical frequencies, intersubband transitions of semiconductor cavities placed in doped potential wells [43-47] form the physically ultrastrong interaction. This is while at the microwave frequencies, superconductor resonator circuits supercooled to milli-Kelvin temperatures in resonance with two-level emitters formed from Josephson junctions may interact ultrastrongly [48,49]. More recently at the Terahertz frequencies, a third ultrastrongly coupled system has been identified [50], which is obtained from the interaction of magnetic cyclotron resonances of a high-mobility two-dimensional electron gas in an amplifying medium. Since the cyclotron frequency, as the transition frequency, is very well controllable with a perpendicular magnetic field, the transition dipoles are easily controlled and may be increased to extremely high values. Rabi frequencies as large as 58% of the transition frequencies have been so far measured [50]. Also, metal-dielectric-metal microcavities along with quantum wells have been shown to form ideal systems for generation of cavity polaritons at the Terahertz spectrum, in which Rabi frequencies exceeding 50% of the transition frequency have been demonstrated [51].

All these three interaction regimes are supposed to be describable by a unified JCPM theory of quantum optics. Recently, we made an attempt to describe the most general CQED system [52] comprising an arbitrary number of emitters and radiation modes subject to an arbitrary initial state. Useful mathematical formulation and analysis of such a system is highly contingent on a different and extended notation of atomic and field states and operators, which we had constructed therein. We furthermore have allowed field-dipole and dipole-dipole interactions to exist. Then the model Hamiltonian was transformed to the Heisenberg's interaction picture under Rotating Wave Approximation (RWA), and the subsequent Rabi equations were numerically solved, as is routinely done elsewhere, too [1].

However, as we are going to discuss it just below, this general approach is mathematically incorrect, especially for the ultrastrongly coupled systems. And, this is *not* because of the RWA, but rather the Heisenberg's transformation involved, which is employed in an incorrect manner. We are going to describe a universal approach, instead, which is mathematically consistent and easy to deal with, and may or may not include RWA. Moreover, our proposed method leads to explicitly closed-form solutions, which may be rapidly evaluated.

In the JCPM, the Hamiltonian is normally transformed into the Heisenberg's interaction picture [1], while it is taken as granted that the Bosonic field creation and annihilation operators should obey simple first-order differential equations with solutions varying in time as $\exp\left(\pm \mathrm{i}\,\omega t\right)$. These *free-running* solutions for field operators clearly oscillate completely sinusoidal in time. In our recent studies [53-56] it has been noticed that these sinusoidal free-running solutions for field operators are not correct, and they in fact oscillate largely non-sinusoidal. Consequently, not only RWA should be avoided for such ultrastrongly coupled systems, but also Heisenberg's transformation must be abandoned.

It is the purpose of this paper to construct and accurately solve a universal, self-consistent, and most general theoretical picture of multipartite CQED systems without RWA and/or Heisenberg's transformation, while being applicable to ultrastrongly coupled systems. The explicit solution is greatly simplified and accurately evaluated using algebraic matrix exponential techniques, as just reported in a recent publication [56]. This technique allows rapid and accurate evaluation of state kets in time without relying on any numerical integration. For treatment of partitions (or more accurately, particles) having spin, and in particular Fermions with half-integer spin, we may note that the corresponding spinors are actually special cases of two-partite entangled systems obtained by outer product of two scalar Fermions. Hence, the present formulation is equally capable of dealing with spin at the expense of a two-fold increase in the number of scalar partitions. As a result, we have been able for the first time to investigate ultrastrongly coupled multipartite systems not been studied so far [57]. These include an ultrastrongly coupled integrated waveguide structure realized in compound semiconductor quantum wells [58], which is modeled as a two-partition system consisting of a three-state $\Lambda$-emitter and one radiation modes. The other example is a seven-partition system consisting of six identical two-level quantum dots and one radiation mode, having a maximum photon occupancy number of twenty-four.

By plotting various probabilities and expectation values of field and atomic operators, we can find that ultrastrongly coupled systems are marked by onset of a chaotic behavior in phase space. Earlier last year [56] we had reported anomalous and largely nonlinear variations for phases of field operators. Our present study for the first time sets up a rigorous and novel method to trace the evolution of multi-partite systems in phase space, in which such chaotic behavior are easily detectable. Our computer software code is theoretically capable of dealing with any multipartite CQED system, and is able to self-generate an internal subroutine for exact calculation of concurrence. This parameter represents the overall degree of entanglement in a multipartite system. Finally, we present a complete numerical simulation of spin entanglement between a photon polarization and electron spin confined in a semiconductor quantum dot. We also demonstrate theoretically how it would be possible to treat spin as an extra degree of freedom in a multi-partite quantum system.

## II. MATHEMATICS AND ALGORITHM

To analyze the general behavior in CQED of complex multi-partite systems, initially the coefficients matrix of the most general system are computed and numerically measured. Such systems normally consist of an arbitrary number of emitters (usually quantum dots) in interaction with an arbitrary number of cavity modes. For this purpose, the general time-dependent state of the most general possible system has been rigorously specified and is solved exactly in time-domain using an explicit analytical solution presented in this section. As it will be shown, providing initial conditions as one of the parts of the solution is vital. Fock and coherent initial conditions are considered in this article, so the most general equation to provide such coherent initial condition is extracted. Required equations to measure the presence probability of the system at different states are presented. Expectation values of field and atomic operators as well as the expectation value of commutator of atomic ladder operators are also extracted. Finally an extensive high-level MATLAB code is developed which sets up the initial conditions for any arbitrary complex system and evaluates mentioned parameters including probabilities, expectation values of field and atomic operators, the commutator of atomic ladder operators, as well as concurrence as the most general measure of entanglement of multipartite systems.

### A. Coefficients Matrix

The aim of the developed code is to solve the Schrödinger equation given by [59]

$$\frac{\partial}{\partial t}\left|\varphi\left(t\right)\right\rangle = -\frac{\mathrm{i}}{\hbar}\mathbb{H}\left|\varphi\left(t\right)\right\rangle \tag{1}$$

where $\left|\varphi\left(t\right)\right\rangle$ is the general state of system, $\mathbb{H}$ is the generalized JCPM Hamiltonian presented in [52] and $\hbar$ is the reduced Plank constant. In Schrödinger space the ket states of the system are time-dependent while operators are not. Due to the reasons discussed earlier [53-56], Heisenberg's transformation must not be used.

We suppose that $|A\rangle$ is the ket state of the different energy levels of emitters, $k$ is the total number of emitters, $|F\rangle$ is the ket state of the cavity modes, $f_\nu$ is the number of photons in $\nu$–th cavity mode number, and $\omega$ is the total number of cavity modes. The ket denoting the $r_n$-th state of light emitting system $\left|{}^{n}_{r_n}\right\rangle$, expresses the condition that the $n$-th quantum dot resides at its $r_n$-th energy level. Now, the general time-dependent state of the most general possible system will be given by

$$\begin{aligned} &|\varphi(t)\rangle = \sum_{A,F} \phi(A,F)|A\rangle|F\rangle \\ &|A\rangle = \bigotimes_{n=1}^{k} \left|{}^{n}_{r_n}\right\rangle = \left|{}^{1}_{r_1}\right\rangle\left|{}^{2}_{r_2}\right\rangle\ldots\left|{}^{k}_{r_k}\right\rangle \qquad 1<r_n<B_n \\ &|F\rangle = \bigotimes_{\nu=1}^{\omega} |f_\nu\rangle = |f_1\rangle|f_2\rangle\ldots|f_\omega\rangle \qquad 0<f_\nu<N_\nu \end{aligned} \tag{2}$$

where $r_n$ refers to the different energy level states of light emitting systems and $B_n$ is the maximum number of energy levels of the $n$-th quantum dot. $N_\nu$ is the number of maximum photons which possibly occupies a cavity mode [52], and here is taken to be identical for all modes. Also, $|\varphi(t)\rangle$ is a superposition of all possible states of the system, including atom and field states, and each state has a time dependent coefficient equal to $\phi(A,F)$.

So according to (2), various states are formed by the different states of quantum dots at different energy levels multiplied by different photon number states in each cavity mode. If $m$ is the total number of cavity modes, then (2) can be written as

$$\left|\varphi(t)\right\rangle = \sum_{r_1,r_2,\ldots,r_n=1}^{B_n} \sum_{f_1,f_2,\ldots,f_\nu,\ldots,f_m=0}^{N} \phi\left(r_1,r_2,\ldots,r_n,f_1,f_2,\ldots,f_\nu,\ldots,f_m\right)\left|A\right\rangle\left|F\right\rangle \tag{3}$$

The generalized JCPM Hamiltonian $\mathbb{H}$ is here consisting of three parts as [52,56]

$$\mathbb{H} = \hat{\mathbb{H}}_0 + \hat{\mathbb{H}}_{\mathbf{r}\cdot\mathbf{E}} + \hat{\mathbb{H}}_{\mathbf{r}\cdot\mathbf{r}} \tag{4}$$

$$\hat{\mathbb{H}}_0 = \sum_{n,i} E_i^n \hat{\sigma}_i^n + \sum_{\nu} \hbar\Omega_\nu \hat{a}_\nu^{\dagger} \hat{a}_\nu \tag{5}$$

$$\hat{\mathbb{H}}_{\mathbf{r}\cdot\mathbf{E}} = \sum_{n,i<j}\left(\gamma_{nij}\hat{\sigma}_{i,j}^n + \gamma_{nij}^{*}\hat{\sigma}_{j,i}^n\right)\sum_{\nu}\left(g_{nij\nu}\hat{a}_\nu + g_{nij\nu}^{*}\hat{a}_\nu^{\dagger}\right) \tag{6}$$

$$\hat{\mathbb{H}}_{\mathbf{r}\cdot\mathbf{r}} = \sum_{n<m,i<j}\left(\eta_{nij}\hat{\sigma}_{i,j}^n + \eta_{nij}^{*}\hat{\sigma}_{j,i}^n\right)\left(\eta_{mij}\hat{\sigma}_{i,j}^m + \eta_{mij}^{*}\hat{\sigma}_{j,i}^m\right) \tag{7}$$

in which $\hat{\mathbb{H}}_0$ describes the system energy without interaction, $\hat{\mathbb{H}}_{\mathbf{r}\cdot\mathbf{E}}$ stands for light-emitter interactions, and $\hat{\mathbb{H}}_{\mathbf{r}\cdot\mathbf{r}}$ represents interactions between any possible pair of emitters such as dipole-dipole terms [52,56]. Coefficients $\gamma_{nij}$ are matrix elements of dipole operator of $n$-th emitter. The strength of the dipole interaction between $n$-th emitter and $\nu$-th mode of cavity is given by $g_{nij\nu}$ with the transition $i$-th and $j$-th energy levels. Coefficients $\eta_{nij}$ are proportional to the strength of the dipole generated while another emitter undergoes a transition between $i$-th and $j$-th levels. $E_i^n$ indicates the $i$-th energy of the $n$-th emitter. Furthermore $\hat{a}_\nu^{\dagger}$ and $\hat{a}_\nu$ are the field creation and annihilation operators which respectively increase and decrease the number of existing photons within the $\upsilon$-th cavity mode by one, $\hat{\sigma}_{s,k}^l$ is atomic ladder operator which makes the $l$-th emitter to switch from $k$-th to $s$-th level, and $\left(\hat{\sigma}_{s,k}^l\right)^{\dagger} = \hat{\sigma}_{k,s}^l$ is its adjoint.

By equalizing the coefficients of similar kets on both sides of (1), the coefficients may be rearranged as elements of a square matrix with dimension $N$ as $\left[M\right]_{N\times N}$ [18] as

$$\frac{\partial}{\partial t}\left\{\varphi\left(t\right)\right\}_{N\times 1}=\left[M\right]_{N\times N}\left\{\varphi\left(t\right)\right\}_{N\times 1} \tag{8}$$

in which $\left\{\varphi\left(t\right)\right\}$ is the vector of unknown coefficients. We suppose that $\left\{\varphi\left(0\right)\right\}$ is the initial condition vector. Then the solution to (8) is simply given by

$$\left\{\varphi\left(t\right)\right\}=\mathrm{e}^{\left[M\right]t}\left\{\varphi\left(0\right)\right\} \tag{9}$$

To evaluate (9), $\left[M\right]$ is first diagonalized into a diagonal matrix $\left[D\right]=\left[D_i\delta_{ij}\right]$ of eigenvalues $D_i$ using the diagonalizer $\left[R\right]$ which is found from eigenvectors of $\left[M\right]$ as

$$\left\{\varphi\left(t\right)\right\}=\mathrm{e}^{\left[M\right]t}\left\{\varphi\left(0\right)\right\}=\left[R\right]\mathrm{e}^{\left[D\right]t}\left[R\right]^{-1}\left\{\varphi\left(0\right)\right\}=\left[R\right]\left[e^{D_i t}\delta_{i,j}\right]\left[R\right]^{-1}\left\{\varphi\left(0\right)\right\} \tag{10}$$

The solution (10) is explicit and can be accurately evaluated regardless of the time $t$. It also excludes the need of matrix exponentiation and is thus numerically stable. We also note that $\left\{\varphi\left(0\right)\right\}$ can take on any initial such as Fock or coherent [60] initial states. For the most general multipartite system consisting of an arbitrary number of modes and emitters, the normalized initial coherent states will be

$$\left\{\varphi\left(0\right)\right\}\left|\varphi\left(t_0\right)\right\rangle=\frac{1}{\sqrt{2^k}}\left\{\prod_{k=1}^{m}\left(\sum_{\alpha_k=0}^{N}\left|\sqrt{\frac{\lambda_k^{\alpha_k}}{\alpha_k!}\mathrm{e}^{-\lambda_k}}\right|^2\right)\right\}^{-\frac{1}{2}}\sum_{k=1}^{m}\sqrt{\prod_{l=1}^{m}\frac{\lambda_l^{n_k}}{n_k!}\mathrm{e}^{-\lambda_l}}\,|\,n_1,\ldots,n_k\rangle\times|\,\text{energy states}\rangle \tag{11}$$

### B. Probabilities of Presence at States

According to (3), the presence probability of an arbitrary light emitting system such as $l$, being in at arbitrary energy level such as $k$ is simply

$$P = \sum_{A-\{r_1\}=1}^{B_n} \sum_{f_1,f_2,\ldots,f_\nu=0}^{\mathrm{N}} \left|\varphi\left(r_1, r_2, \ldots, r_{l\to k}, r_n, f_1, \ldots f_\nu\right)\right|^2 \tag{12}$$

### C. Expectation Values of Field Operators

Expectation values of annihilation operator for the most general possible complex system are found as

$$\left\langle \varphi(t)\right| \hat{a}_\nu \left|\varphi(t)\right\rangle = \sum_{A,F} \sqrt{f_\nu}\,\phi^*\left(A, f_\nu - 1\right)\ \phi\left(A, f_\nu\right)$$
$$= \sum_{r_1,r_2,\ldots,r_n=1}^{B_n} \sum_{f_1,f_2,\ldots,f_\nu,\ldots,f_m=0}^{N} \sqrt{f_\nu}\,\phi^*\left(r_1, r_2, \ldots, r_n, f_1, f_2, \ldots, f_\nu - 1 \ldots, f_m\right)\phi\left(r_1, r_2, \ldots, r_n, f_1, f_2, \ldots, f_\nu, \ldots, f_m\right)$$

(13)

### D. Expectation Values of Ladder Operators

Expectation values of the atomic transition operator are found as

$$\left\langle \varphi(t)\right| \hat{\sigma}^l_{s,k} \left|\varphi(t)\right\rangle = \sum_{A-\{r_l\},F} \phi^*\left(A_{r_l\to s}, F\right)\left\langle A_{r_l\to s}\right|\left\langle F\right| \times \phi\left(A_{r_l\to k}, F\right)\left|A_{r_l\to k}\right\rangle\left|F\right\rangle$$
$$= \sum_{A-\{r_l\},F} \phi^*\left(A_{r_l\to s}, F\right)\phi\left(A_{r_l\to k}, F\right)$$
$$= \sum_{A-\{r_l\}=1}^{B_n} \sum_{f_1,f_2,\ldots,f_m=0}^{N} \phi^*\left(r_1, r_2, r_l \to s \ldots, r_n, f_1, f_2, \ldots, f_m\right)\phi\left(r_1, r_2, r_s \to k, \ldots, r_n, f_1, f_2, \ldots, f_\nu, \ldots, f_m\right)$$

(14)

### E. Expectation Value of the Ladder Commutator

The commutator of the atomic ladder operators is the commutation of atomic transition operator and its Hermitian adjoint, denoted by $\left[\hat{\sigma}^l_{s,k}, \left(\hat{\sigma}^l_{s,k}\right)^\dagger\right]$, with the expectation value found as

$$\left\langle \varphi(t)\right|\left[\hat{\sigma}^l_{s,k}, \left(\hat{\sigma}^l_{s,k}\right)^\dagger\right]\left|\varphi(t)\right\rangle = \sum_{A-\{r_l\},F}\left|\phi^*\left(A_{r_l\to s}, F\right)\right|^2 - \sum_{A-\{r_l\},F}\left|\phi^*\left(A_{r_l\to k}, F\right)\right|^2 \tag{15}$$

## III. ANALYSIS AND NUMERICAL RESULTS

Here, we present detailed analysis of two different quantum optical systems: a three-level quantum well and a multipartite quantum optical system with seven partitions. All extracted equations in the previous section by the utilization of our provided code are executed. Entanglement is also analyzed in both systems.

### A. CQED in an Optoelectronic Device

In this section, we present the simulation and analysis of the CQED of a real complex system consisting of a three level light emitting system interacting with a cavity mode. The emitter is an InGaAlAs quantum well and the light is guided in a waveguide underneath [57]. The design details and applications of such optoelectronic device as a wide-band and ultra-compact optical modulator is discussed elsewhere [57]. It has been shown that for the system of interest, the quantum well could be modeled as a three-level light emitting system with defined energy levels, corresponding to electrons, and heavy and light holes bands as in Fig. 1. Transition dipole moments between different energy levels are also calculated in [57].

#### *1. System characteristics*

The light emitting system described in our other article [57] is a heterostructure $In_{0.52}(Al_xGa_{1-x})As/In_{0.53}Ga_{0.47}As/In_{0.52}(Al_xGa_{1-x})_{0.48}As$ quantum potential well. For an Aluminum fraction of $x$=0.9 and well material thickness of 9nm, the transition energy between conduction and heavy hole bands will be about 0.8eV. There is a 30meV offset between the heavy and light holes bands. We thus may choose the energy levels of the model light emitter system respectively as 0, 30, and 829 meV for light holes, heavy holes, and electrons bands. Rabi frequencies are calculated as

$$G_h = \frac{1}{\hbar}(E_0\hat{E})\left\langle \psi_e | e\mathcal{R} | \psi_{hh} \right\rangle \tag{16}$$

$$G_l = \frac{1}{\hbar}(E_0\hat{E})\left\langle \psi_e | e\mathcal{R} | \psi_{lh} \right\rangle \tag{17}$$

in which $G_h$ and $G_l$ are the Rabi frequencies for transition between conduction band to heavy holes and light holes, respectively. $\left\langle \psi_e | e\mathcal{R} | \psi_{hh} \right\rangle$ and $\left\langle \psi_e | e\mathcal{R} | \psi_{lh} \right\rangle$ are transition dipole moments, respectively calculated as 26.15 Debye and 15.21 Debye [57], and $E_0\hat{E}$ is the applied electrical field.

TABLE I. Rabi frequencies and Coupling regimes.

| Coupling Regime | $E_0$ | $G_h\,(\frac{\text{Rad}}{\text{s}})$ | $G_l\,(\frac{\text{Rad}}{\text{s}})$ | $\frac{G_h}{\omega_\lambda}$ | $\frac{G_l}{\omega_\lambda}$ |
|---|---|---|---|---|---|
| Weak | $100\frac{\text{kV}}{\text{cm}}$ | $8.2640\times10^{12}$ | $4.8067\times10^{12}$ | 0.0066 | 0.0038 |
| Strong | $1\frac{\text{MV}}{\text{cm}}$ | $8.2640\times10^{13}$ | $4.8067\times10^{13}$ | 0.066 | 0.038 |
| Ultrastrong | $10\frac{\text{MV}}{\text{cm}}$ | $8.2640\times10^{14}$ | $4.8067\times10^{14}$ | 0.66 | 0.38 |

Table I shows the range of Rabi frequencies measured based on (16) and (17) for various electric field strengths. It was assumed that the light emitting system is in interaction with photons having the wavelength of $\lambda = 1.49\mu\text{m}$ so the optical frequency $\omega_\lambda$ would be equal to $1.2582\times10^{15}\,\frac{\text{Rad}}{\text{s}}$.

To simulate different coupling regimes,, electrical field is allowed to vary while the features of the considered system are fixed. According to the Table I, and because of near-resonance situation between light and electron-heavy hole transition, in the weakest applied electric field with $E_0 = 100$ kV/cm, the system is in the weak coupling regime. For a stronger electric field with $E_0 = 1$ MV/cm, the coupling regime is strong. Finally, for the strongest electric field with $E_0 = 10$ MV/cm, the coupling enters the ultrastrong regime.

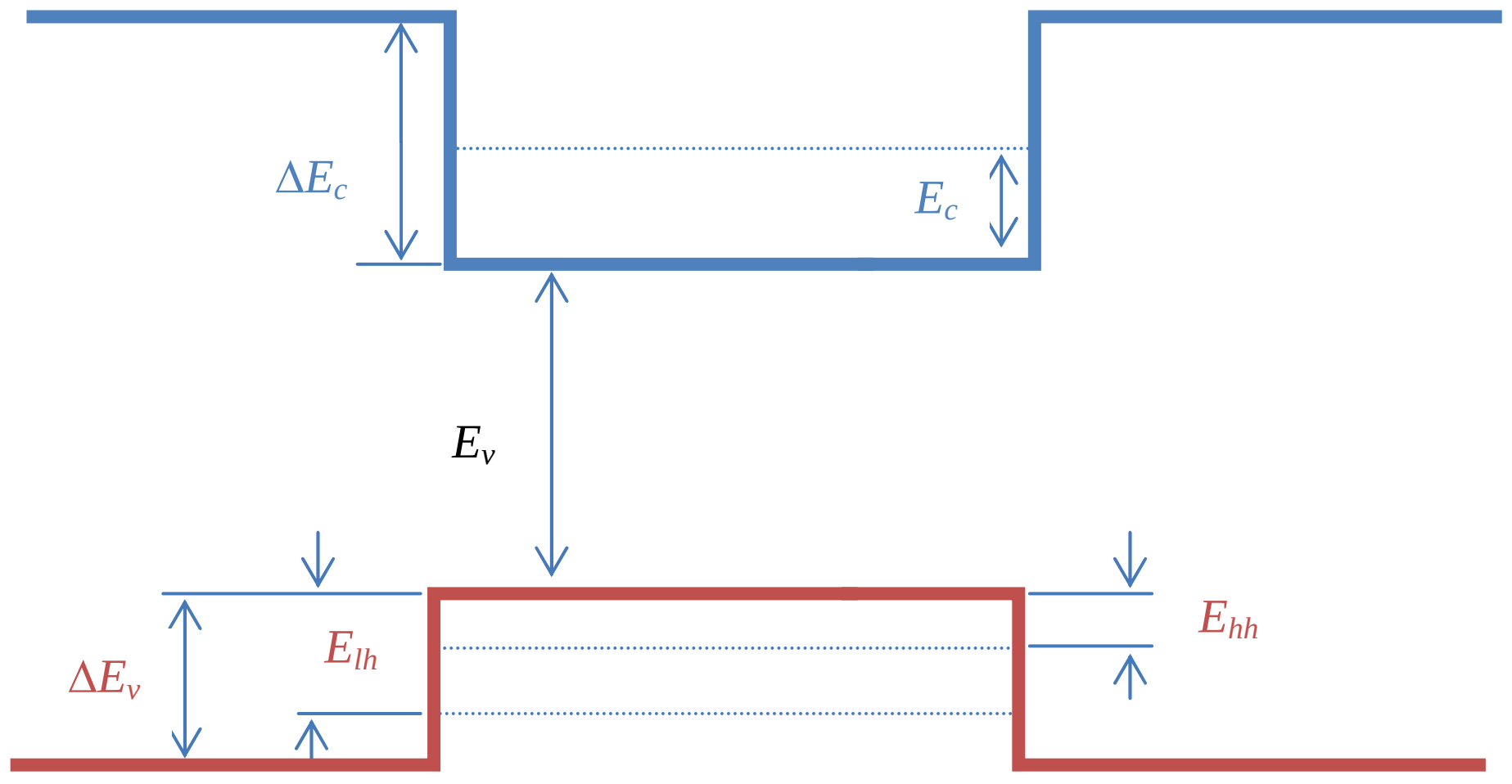

FIG. 1. QW with the electron, heavy-hole and light-hole levels [57].

The general time-dependent state of the system and its describing Hamiltonian according to the Eqs. (2,5,6,7) are now expressed as

$$\left|\varphi(t)\right\rangle = \sum_{A=lh,hh,e,F} \phi(A,F)\left|A\right\rangle\left|F\right\rangle \tag{18}$$

$$\hat{\mathbb{H}}_0 = \sum_{1,i} E_i^1 \hat{\sigma}_i^1 + \hbar\Omega \hat{a}^\dagger \hat{a}$$
$$\hat{\mathbb{H}}_{\mathbf{r}\cdot\mathbf{E}} = \sum_{1,i<j} \left( \gamma_{nij} \hat{\sigma}_{i,j}^1 + \gamma_{nij}^* \hat{\sigma}_{j,i}^1 \right) \sum_{\nu} \left( g_{nij} \hat{a} + g_{nij}^* \hat{a}^\dagger \right) \tag{19}$$

In this system, $\hat{\mathbb{H}}_{\mathbf{r}\cdot\mathbf{r}}$ is equal to zero because there is only one emitter. As input of our software all coupling coefficients are entered in units of energy. The coherent initial state in this system follows (11).

*2. Presence probabilities with Fock initial state*

By applying Fock initial conditions we intended to study two cases. Firstly, to study the effect of boosting coupling coefficient on the presence probability of the system in normalized time, secondly, to study the importance of coefficient matrix measurement of the system exactly and without RWA. As we know, in RWA the effect of two terms $\left(\hat{\sigma}_{i,j}^1\right)^\dagger \hat{a}^\dagger = \hat{\sigma}_{j,i}^1 \hat{a}^\dagger$ and $\hat{\sigma}_{i,j}^1 \hat{a}_\nu$ are neglected.

Since the light emitting system under consideration consists of three energy levels, the effect of both $\hat{\sigma}_{e,lh} \hat{a}^\dagger$ and $\hat{\sigma}_{e,hh} \hat{a}^\dagger$ could be studied. To study the $\hat{\sigma}_{e,lh} \hat{a}^\dagger$ term due to Fock initial condition, $|1,lh\rangle$ state is considered. For instance under the operation of $\hat{\sigma}_{e,lh} \hat{a}^\dagger$ the state ket will become

$$\sqrt{2}\gamma_{1,lh,hh} g_{1,lh,hh}^* |2,hh\rangle + \sqrt{2}\gamma_{1,lh,e}^* g_{1,lh,e}^* |2,e\rangle \tag{20}$$

Since, heavy to light holes transition is forbidden, we have $g_{1,lh,hh} = 0$, and hence only the probability presence of the system in the $|2,e\rangle$ state should be calculated, which is plotted for $|2,e\rangle$ and $|1,lh\rangle$ as a function of normalized time different coupling regimes in Figs. 2-3. The effect of $\hat{\sigma}_{e,hh}\hat{a}^{\dagger}$ term may be studied similarly.

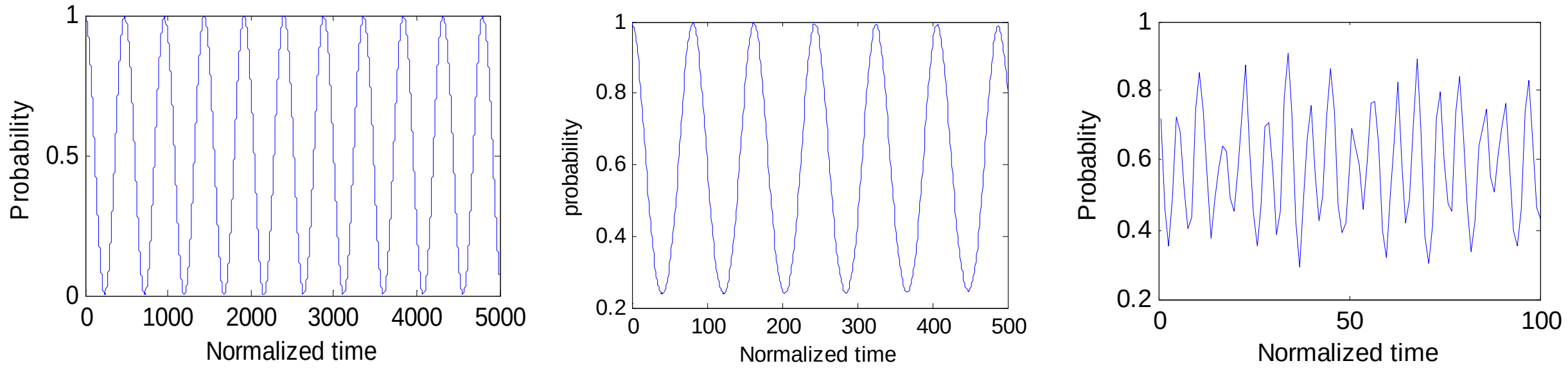

FIG. 2. The presence probability of the system in $|1,lh\rangle$ under weak, strong and ultrastrong coupling, respectively from left to right.

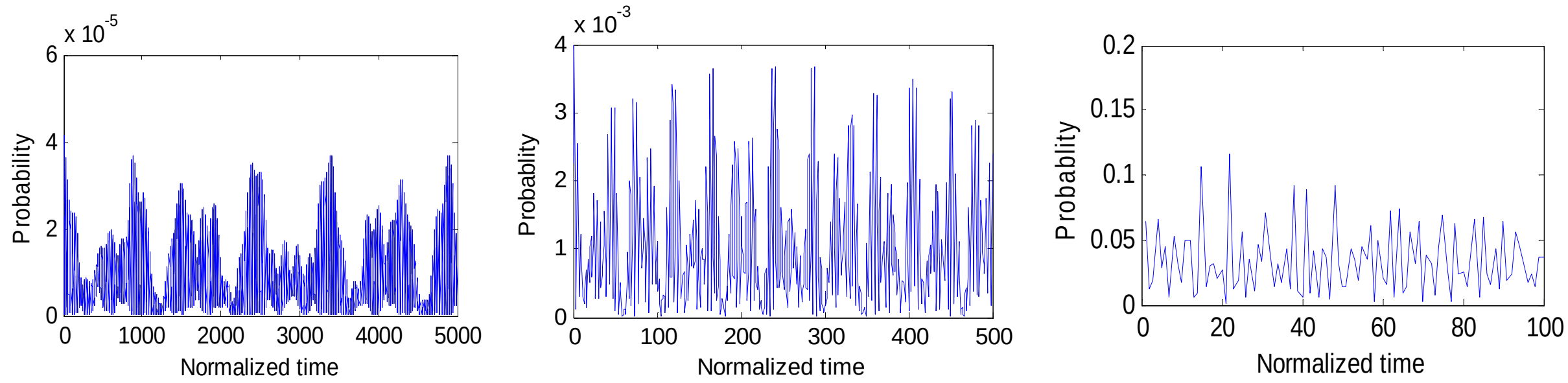

FIG. 3. The presence probability of the system in $|2,e\rangle$ under weak, strong and ultrastrong coupling regimes respectively from left to right.

As it is here seen, by increasing coupling strength, the frequency of oscillations also increases, except for the ultrastrong regime which exhibits a non-sinusoidal and disordered behavior or anonymous oscillations. It is also observed that although the probability of the states which are neglected in RWA, is negligibly small in weak and strong coupling, while it is significantly larger in the ultrastrong coupling. Hence, RWA is a very inappropriate approximation for study of ultrastrong coupling [62].

*3. Presence probabilities with coherent initial state*

Following (12), the presence probability of system at each of the conduction, heavy hole, or light hole levels. This probability is also found and plotted in the Fig. 4. By comparison it is observed that in the weak and strong coupling regimes, the presence probability of the system in different states is sinusoidal on short time scales. The probability for a maximum photon occupancy number of 8 is

$$P = \sum_{f_1=0}^{8} \left| \varphi\left( r_{1\rightarrow e}, f_1 \right) \right|^2 \tag{21}$$

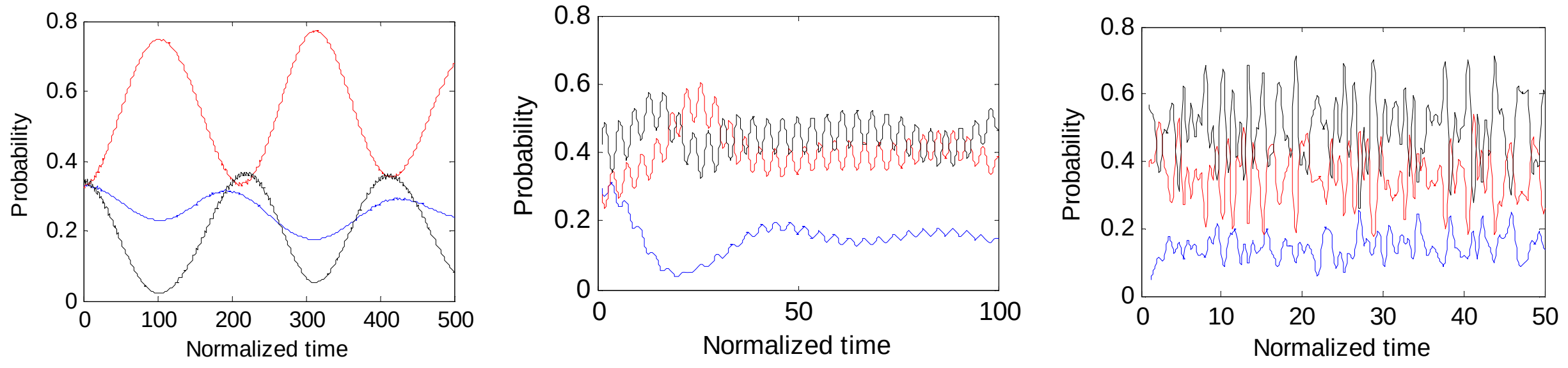

FIG. 4. Probabilities of occupation of the, light-hole (lh), heavy-hole (hh) and conduction (e) states (blue, red and black in each figure respectively); from left to right: weakly, strongly, and ultrastrongly coupled systems.

This is while by entering into ultracoupling regimes, the behavior is not sinusoidal and has anonymous oscillations at all on any time scale and is very chaotic both in short and long temporal range. Remarkably, nonlinear quantum optical chaos in an optical beam has been observed recently in experiment [61] through direct measurement of the evolution of Wigner function in phase space.

*4. Annihilation in different coupling regimes*

Following (13), the expectation value of the field annihilation operator of the system having coherent initial state is measured. In order to study the behavior of field operators in different coupling regimes, this expression has been calculated. Since the annihilator is non-Hermitian, its expectation value is complex-valued. Making a three-dimensional parametric plot having the corresponding real and imaginary parts as functions of normalized time is very instructive in this case. This has been shown for various coupling regimes and shown in Figs. 5-6. Also the phase of expectation value as a function of normalized time duration has been also plotted in separate diagrams therein.

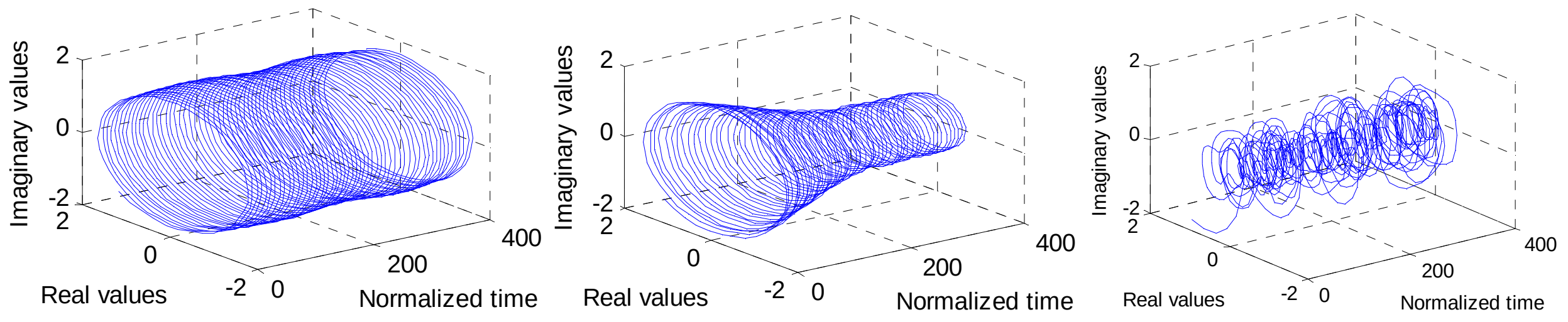

FIG. 5. Three-dimensional plots of the real and imaginary values of the expectation value of $\hat{a}$ in weakly, strongly and ultrastrongly coupled system from left to right.

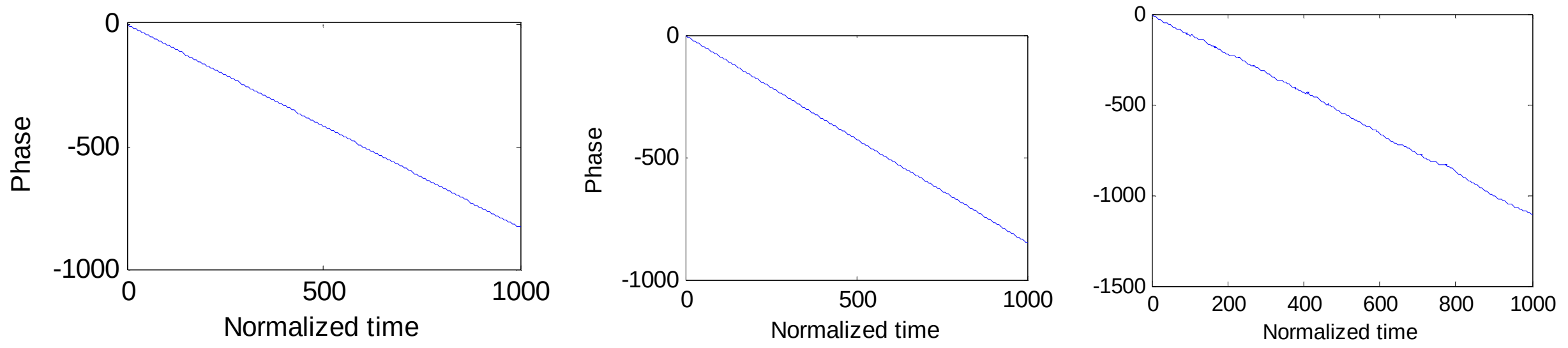

FIG. 6. Phase plots of the real and imaginary values of the expectation value of $\hat{a}$ in weakly, strongly and ultrastrongly coupled systems from left to right.

It is observed that in the weak coupling regime the behavior of the expectation value of the field annihilation operator is completely sinusoidal and it confirms that solving Schrödinger equation in Heisenberg space with RWA can be acceptable in this regime. By increasing the coupling constant and entering into strong regime, as it is seen in Fig. 5, the behavior keeps varying sinusoidally, but the amplitude of oscillations gradually decrease with time. This is while the phase behaves just similarly to the weak coupling regime. For the ultrastrongly coupled system, as shown in Fig. 5, however, the expectation value plot is remarkably non-sinusoidal and chaotic and has anonymous oscillations. At the same time, the phase is significantly non-linear. We had also noticed this particular behavior of the phase under ultrastrong coupling in our recent studies [53-56].

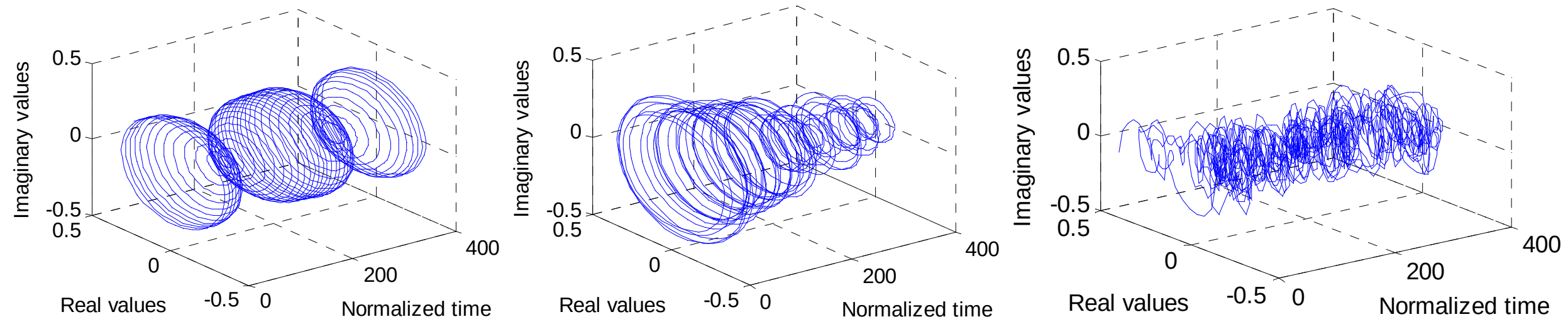

FIG. 7. Three-dimensional plots of the real and imaginary values of the expectation value of $\hat{\sigma}_{hh,e}$ in weakly, strongly and ultrastrongly coupled system from left to right.

### *5. Expectation value of the ladder operator*

According to (14), the expectation value of the atomic ladder operator for transitions between conduction and heavy and light holes while initial state is coherent are measured. These have been calculated and plotted in Figs. 7-10, as parametric plots similar to those of annihilator operator in the above. Similarly, as the system enters the ultrastrong coupling regime, chaotic behavior starts to develop, which is clearly visible both in the phase space and phase plot.

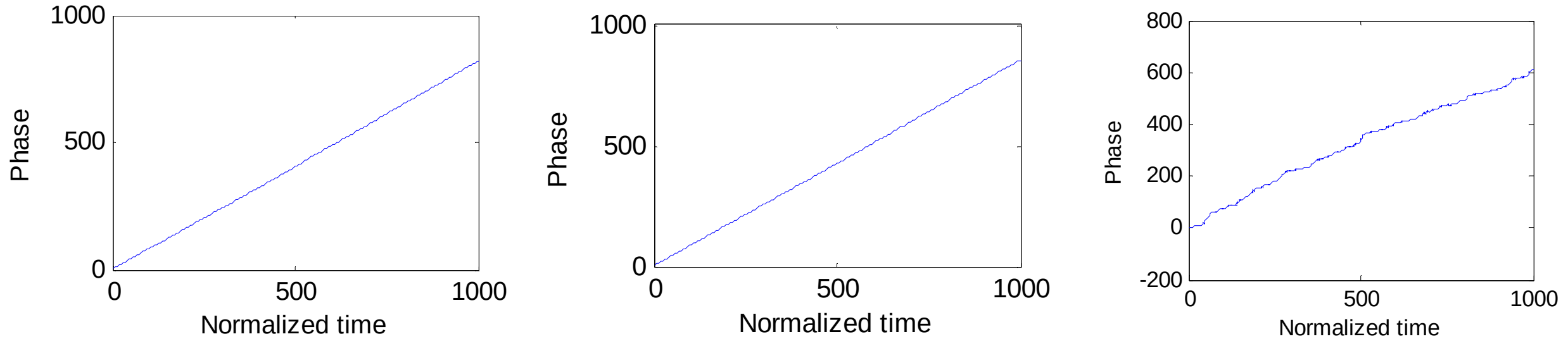

FIG. 8. Phase plots of the real and imaginary values of the expectation value of $\hat{\sigma}_{e,hh}$ in weakly, strongly and ultrastrongly coupled systems from left to right.

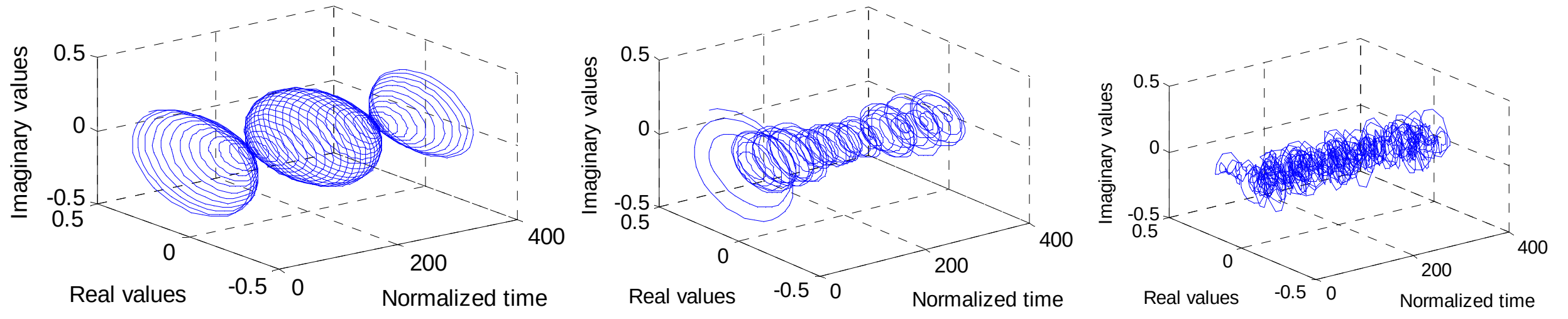

FIG. 9. Three-dimensional plots of the real and imaginary values of the expectation value of $\hat{\sigma}_{lh,e}$ in weakly, strongly and ultrastrongly coupled systems from left to right.

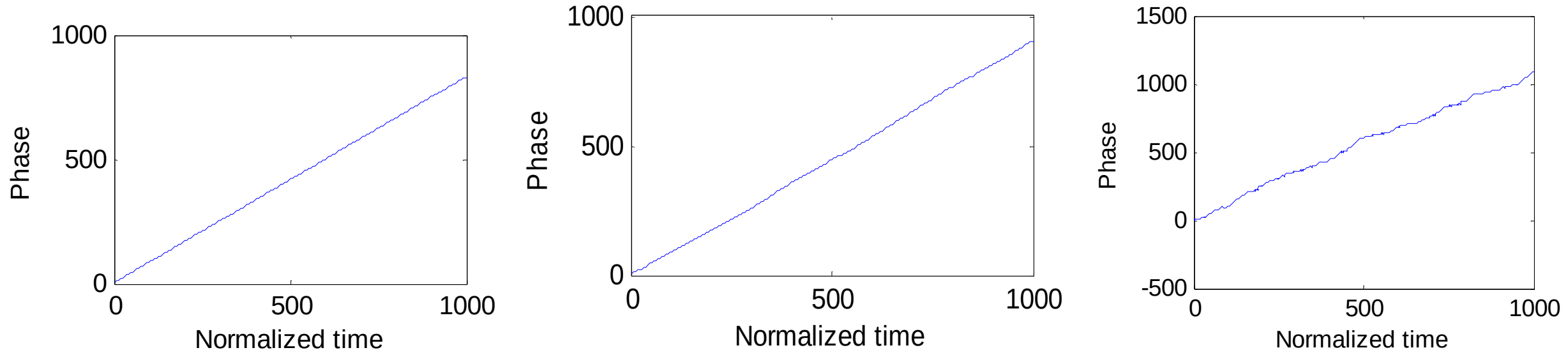

FIG. 10. Phase plots of the real and imaginary values of the expectation value of $\hat{\sigma}_{lh,e}$ in weakly, strongly and ultrastrongly coupled systems from left to right.

*6. Entanglement*

The expectation value of the commutator of atomic ladder operators is known for simple quantum systems to encompass basic information regarding the degree of entanglement. For multipartite systems, however, a more complicated measure such as concurrence [63] should be computed. For the system under study, we calculate and plot both. For an initial coherent state, the expectation of commutator of the ladder operators (15) has been calculated and plotted for various transitions in Figs. 11 and 12, respectively, for hh-e, and lh-e transitions. Again the general trend is such that the strongly coupled system exhibits fast and slow components multiplied together, while this differentiation of fast and slow oscillations in the ultrastrongly coupled systems is not possible.

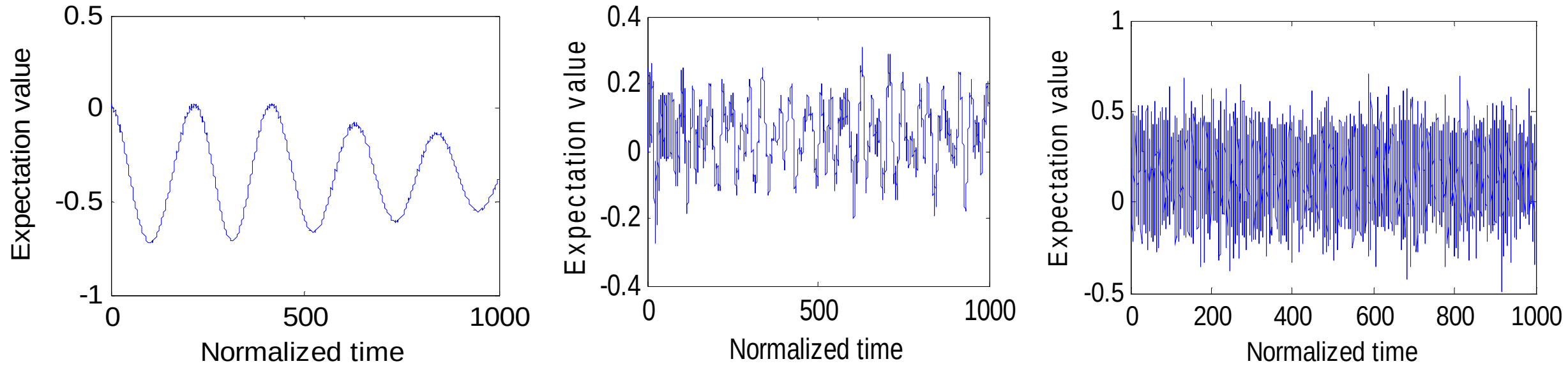

FIG. 11. Expectation value of $\left[\hat{\sigma}_{hh,e},\left(\hat{\sigma}_{hh,e}\right)^{\dagger}\right]$ in weakly, strongly, and ultrastrongly coupled systems from left to right.

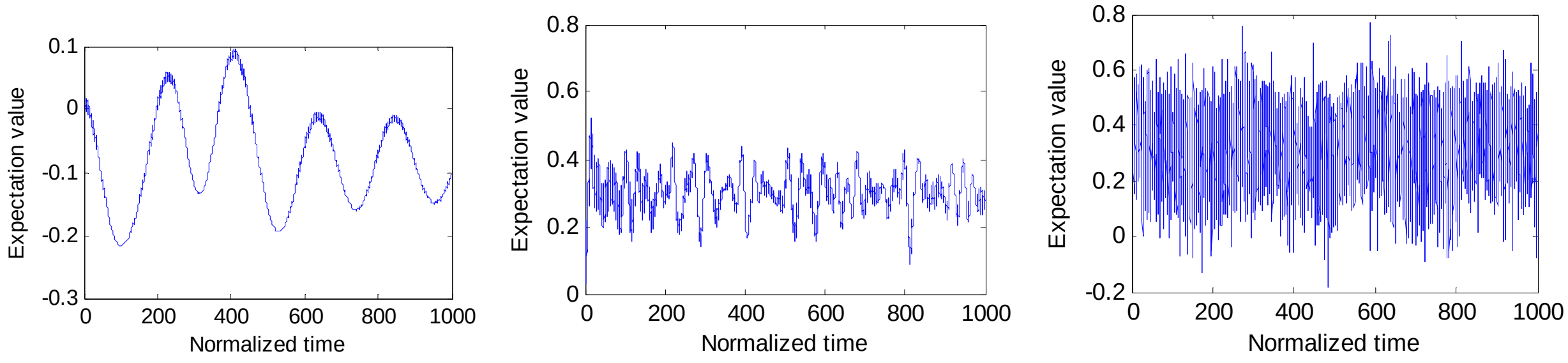

FIG. 12. Expectation value of $\left[\hat{\sigma}_{lh,e},\left(\hat{\sigma}_{lh,e}\right)^{\dagger}\right]$ in weakly, strongly, and ultrastrongly coupled systems from left to right.

By recycling the computer software provided in our previous research [56], which self-generates an internal subroutine for exact calculation of concurrence, we were also able to compute graphs of concurrence under various coupling strengths, ranging from weak to strong and ultrastrong regimes. This has been shown in Fig. 13.

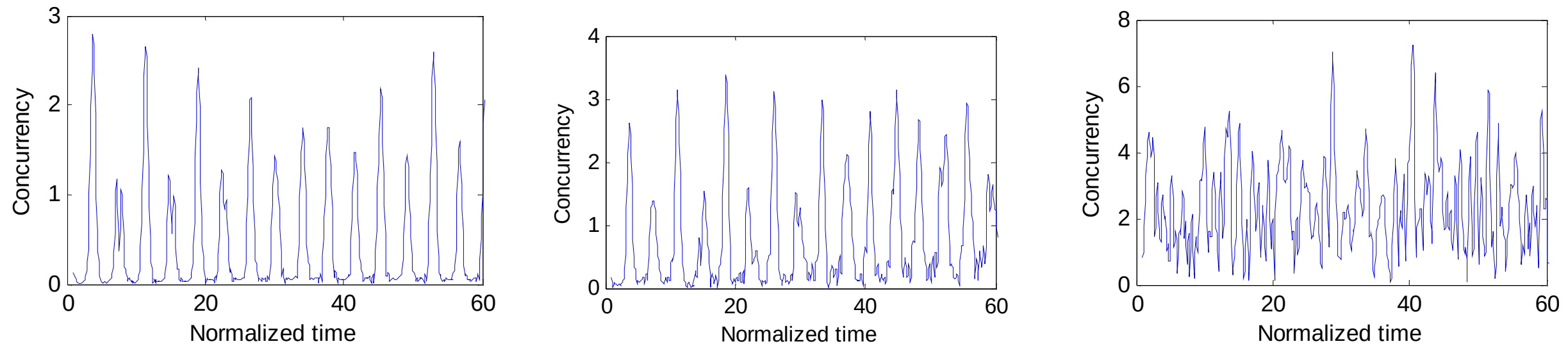

FIG. 13. Computed graphs of concurrency, from left to right: weakly, strongly, and ultrastrongly coupled systems.

It is here concluded through observations made on many multipartite quantum systems that the entanglement of in all cases behaves also chaotic in the ultrastrong regime and exhibits lots of disordered oscillations and distortion. Further details of calculations and plots as well as program source codes are presented for this system and other examples elsewhere [58].

## B. CQED in a seven-partition system

In this section we report the simulation and analysis of a seven-partition system consisting of identical six quantum dots interacting with one cavity mode. Due to the larger number of system partitions, the number of different states of the system increases very much.

*1. System specifications*

The six quantum dots are here limited to two ground and excited states each, with energy eigenvalues of 0eV and 1eV, respectively. The condition on identically of dots may not be achieved in practice, and the developed software code is able to treat different emitters with equal efficiency. The condition is set here for simplification of the problem and reduction of the too many degrees of freedom. It is furthermore supposed that transition dipole moment in these quantum dots is 192 Debye. We also allow mutual dipole-dipole interactions between all the dots with a magnitude of 5meV. It is also assumed that quantum dots are interacting with one resonant cavity mode, having the frequency

$$\omega_\lambda = \frac{E_i - E_g}{\hbar} = \frac{1\,\text{eV}}{\hbar} = 1.5177 \times 10^{15}\,\frac{\text{Rad}}{\text{s}} \tag{22}$$

By applying different electrical fields and comparing with the optical frequency $\omega_\lambda$, different coupling regimes are simulated. According to (16,17), Rabi frequencies as the coupling constants are here calculated and enlisted in Table II, for weakly, strongly, and ultrastrongly coupled systems.

TABLE II. Rabi frequencies and Coupling regimes.

| Coupling Regime | $E_0$ | $G_{e,g}$ $(\frac{\text{Rad}}{\text{s}})$ | $\frac{G_{e,g}}{\omega_\lambda}$ |
|---|---|---|---|
| Weak | $10\frac{kV}{cm}$ | $6.0676\times10^{12}$ | 0.004 |
| Strong | $100\frac{kV}{cm}$ | $6.0676\times10^{13}$ | 0.04 |
| Ultrastrong | $1\frac{MV}{cm}$ | $6.0676\times10^{14}$ | 0.4 |

With the assumption that the maximum possible number of photons in the cavity mode is 8, the general time dependent state of system and the describing Hamiltonian by (2,5,6,7) were specified. Initial Fock and coherent conditions (11) were considered to simulate the system.

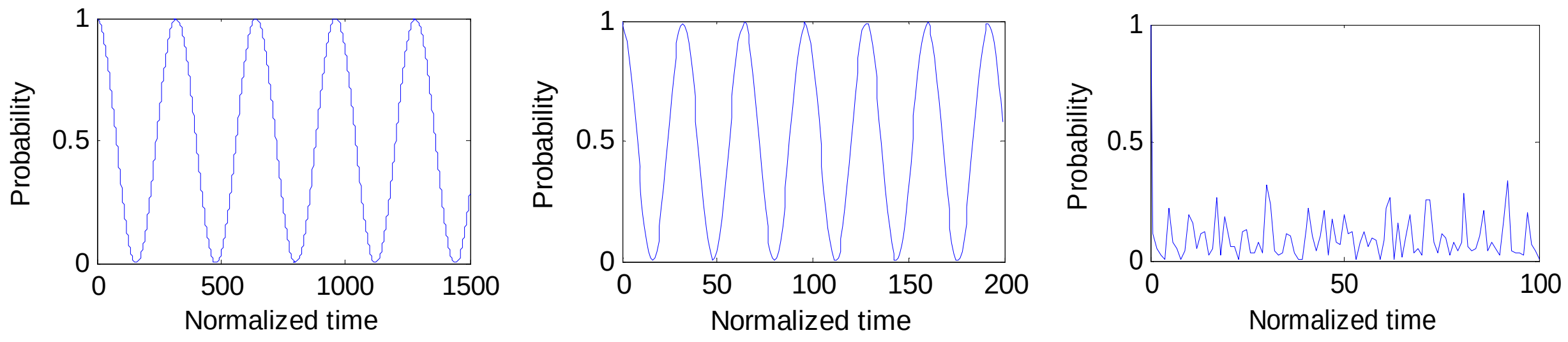

FIG. 14. The presence probability of the system in $|1,g,g,g,g,g,g\rangle$ ket state under weak, strong and ultrastrong coupling regimes from left to right and below respectively.

*2. Presence probabilities with Fock initial state*

We first assume that the initial state is simply $|1,g,g,g,g,g,g\rangle$, which expresses that there is exactly one photon in the cavity mode and all quantum dots are in their ground energy level. Setting this ket as the initial state, we calculate and plot the presence probability in this state as a function of normalized time in different coupling regimes in Fig. 14.

As it is seen in these plots, by increasing the coupling constant the oscillation frequency increases, while in the ultrastrong regime the behavior is chaotic and undergoes anonymous oscillations. This characteristic behavior of the ultrastrong coupling is also justified similarly in the rest of simulations, as discussed in the following.

### *3. Presence probability with coherent initial state*

The presence probability of the first dot being in its ground, or excited energy levels is given according to (12) by

$$
\begin{aligned}
P_g &= \sum_{A-\{r_1\}=g} \sum_{f_{1\cdots 8}=0}^{8} \left|\varphi\left(r_{1\to g}, r_2, \ldots, r_8, f_1\right)\right|^2 \\
P_e &= \sum_{A-\{r_1\}=e} \sum_{f_{1,}=0}^{8} \left|\varphi\left(r_{1\to e}, r_2, \ldots, r_8, f_1\right)\right|^2
\end{aligned}
\tag{23}
$$

Similar expressions may be obviously written for each of the quantum dots. These probabilities for the sixth dot have been plotted in Fig. 15, for weak, strong, and ultrastrong coupling. The characteristic chaotic behavior of ultrastrong coupling can be seen again.

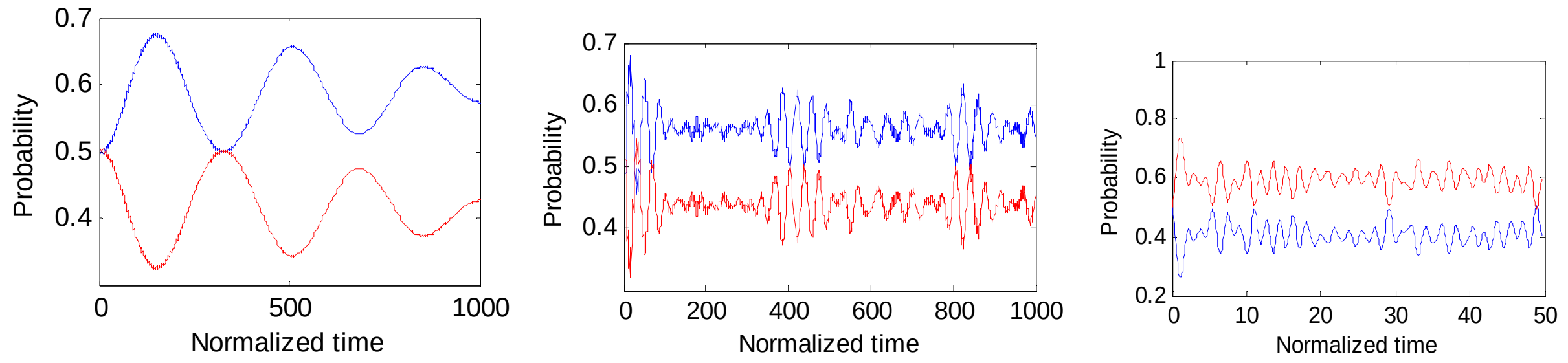

FIG. 15. Probabilities of occupation of the excited and ground energy level states in the sixth quantum dot (red and blue colors in each figure respectively); from left to right: weakly, strongly and ultrastrongly coupled system.

*4. Annihilation in different coupling regimes*

The expectation value of the field annihilation operator of the system is measured based on (13), due to initial coherent state. Phase space and phase plots of the real and imaginary values of the expectation value as functions of normalized time are shown in Figs. 16-17, respectively for weakly, strongly, and ultrastrongly coupled systems.

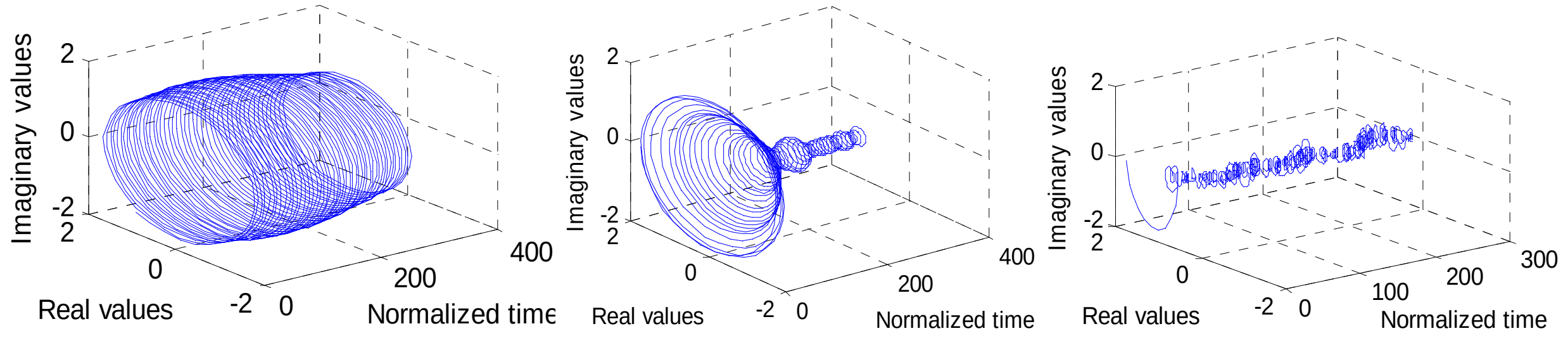

FIG. 16. Three-dimensional plots of the real and imaginary values of the expectation value of $\hat{a}$ in weakly, strongly and ultrastrongly coupled system from left to right.

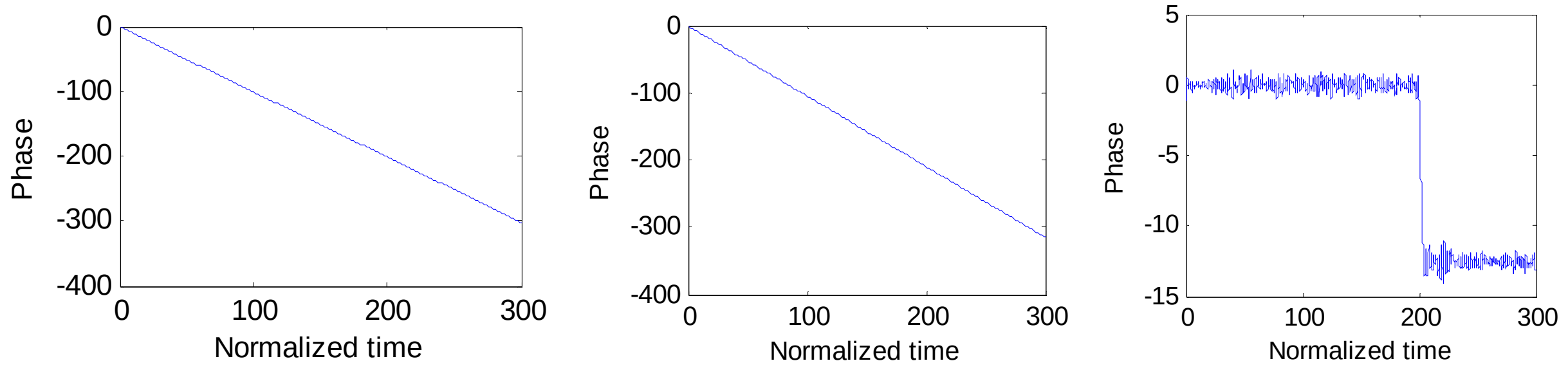

FIG. 17. Phase plots of the real and imaginary values of the expectation value of $\hat{a}$ in weakly, strongly and ultrastrongly coupled systems from left to right.

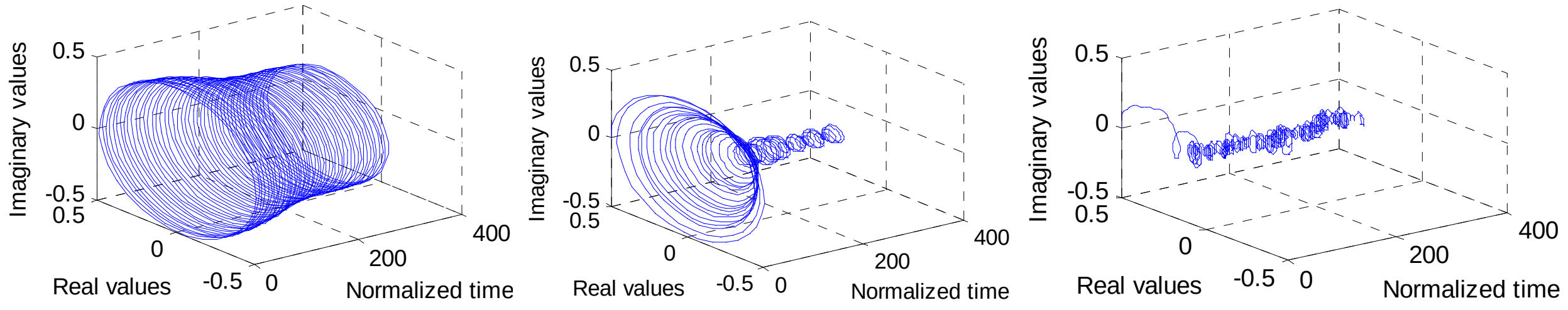

FIG. 18. Three-dimensional plots of the real and imaginary values the expectation value of $\hat{\sigma}$ for the sixth quantum dot in weakly, strongly and ultrastrongly coupled system from left to right.

It is surprisingly observed that by the increasing the coupling constant the entering into ultrastrong regime the coupled system not only exhibits a very chaotic and disordered behavior in the three-dimensional parametric plot, but also the corresponding phase changes abruptly. This behavior is also seen in nearly all other complex expectation values of all ultrastrongly coupled multipartite systems we have studied so far, and is yet to be understood.

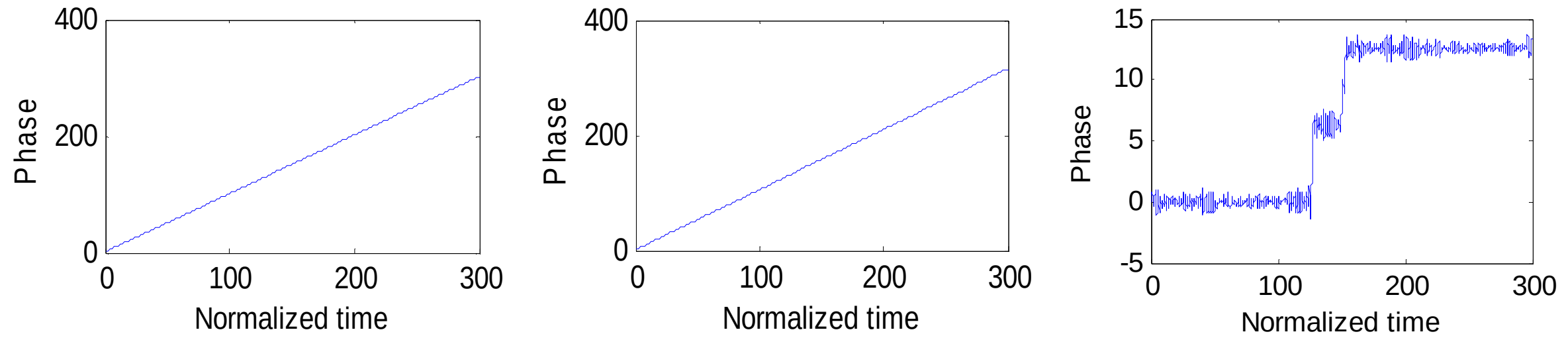

FIG. 19. Phase plots of the real and imaginary values of the expectation value of $\hat{\sigma}$ for the sixth quantum dot in weakly, strongly and ultrastrongly coupled system from left to right.

*5. Expectation value of the atomic ladder operator*

Again we choose the sixth dot as the illustrative example. According to (14), the expectation value of the atomic ladder operator for decay of every quantum dot individually is measured while initial state is coherent. Three-dimensional (phase space) and phase plot of the real and imaginary values of this expectation value as functions of normalized time duration has been similarly plotted in Figs. 18-19. As it is observed in numerical simulations, all six dots behave more or less according to the same pattern with slight differences are seen in the whole system.

All six quantum dots have the same sinusoidal or nearly-sinusoidal oscillations in weakly and strongly coupled systems, respectively. This is while the ultrastrong coupling is accompanied with chaotic three-dimensional trajectories and multi-step random-like abrupt phase changes for all six quantum dots. These abrupt phase changes may find applications in multi-state quantum information processing later, if understood and predicted correctly.

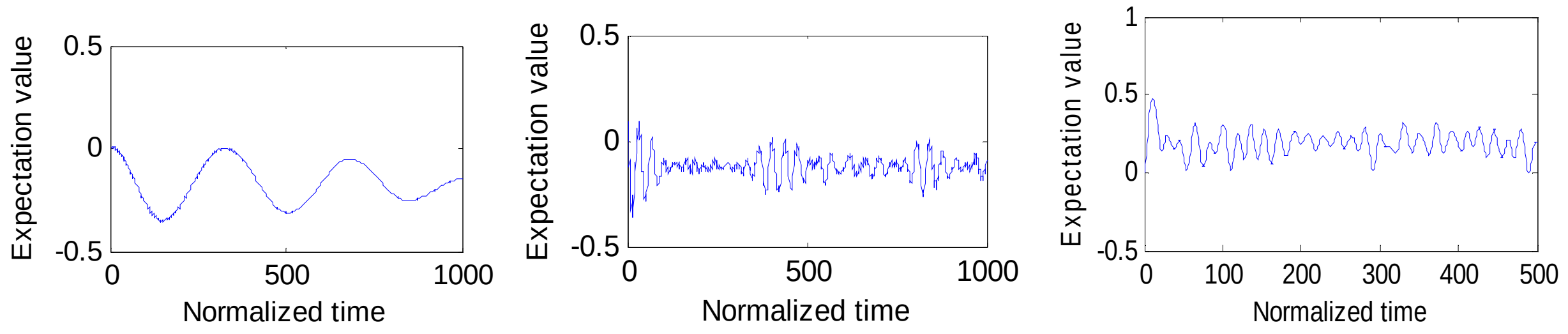

FIG. 20. Expectation value of $\left[\hat{\sigma}_{g,e},\left(\hat{\sigma}_{g,e}\right)^{\dagger}\right]$ in weakly, strongly, and ultrastrongly coupled systems from left to right.

*6. Entanglement*

The expectation value of the commutator of atomic ladder operators for transition of every quantum dot individually from excited energy level to ground energy level has been also analyzed as given by (15). Plots are presented in Fig. 20. It is observed and concluded by the measurements that the system entanglement in all quantum dots is chaotic under the ultrastrong regime and accompanied by a lot of distortion.

## IV. SPIN ENTANGLEMENT

In this section, we offer a method to deal with the systems arising from spin interactions. Such processes including spin-photon entanglement have just recently been observed [64,65]. In order to treat the effect of spin, we rewrite the system state and base kets as

$$
\begin{aligned}
&\left|\varphi(t)\right\rangle = \sum_{A,F} \phi\left(A,F\right)\left|A\uparrow\right\rangle\left|A\downarrow\right\rangle\left|F\uparrow\right\rangle\left|F\downarrow\right\rangle \\
&\left|A\uparrow\right\rangle = \bigotimes_{n=1}^{k} \left|{}^{n}_{r_n}\uparrow\right\rangle = \left|{}^{1}_{r_1}\uparrow\right\rangle\left|{}^{2}_{r_2}\uparrow\right\rangle\ldots\left|{}^{k}_{r_k}\uparrow\right\rangle \qquad 1<r_n<B_n \\
&\left|A\downarrow\right\rangle = \bigotimes_{n=1}^{k} \left|{}^{n}_{r_n}\downarrow\right\rangle = \left|{}^{1}_{r_1}\downarrow\right\rangle\left|{}^{2}_{r_2}\downarrow\right\rangle\ldots\left|{}^{k}_{r_k}\downarrow\right\rangle \\
&\left|F\uparrow\right\rangle = \bigotimes_{\nu=1}^{\omega} \left|f_\nu\uparrow\right\rangle = \left|f_1\uparrow\right\rangle\left|f_2\uparrow\right\rangle\ldots\left|f_\omega\uparrow\right\rangle \qquad 0<f_\nu<N \\
&\left|F\downarrow\right\rangle = \bigotimes_{\nu=1}^{\omega} \left|f_\nu\downarrow\right\rangle = \left|f_1\downarrow\right\rangle\left|f_2\downarrow\right\rangle\ldots\left|f_\omega\downarrow\right\rangle
\end{aligned} \tag{24}
$$

In (24) $\left|A\uparrow\right\rangle$ and $\left|A\downarrow\right\rangle$ denotes the ket states of the different energy levels of the two systems being formed from the ket state, $\left|A\right\rangle$ due to the presence of an external magnetic field. So according to (24), we allow all emitters indexed by $n$ to take on up or down spin states. Similarly, cavity photon Fock states indexed by $\nu$ can take on up or down states (corresponding actually to the two counter-rotating circular polarizations of photons), $\left|F\uparrow\right\rangle$ and $\left|F\downarrow\right\rangle$. Other parameters are those introduced in (2). Expansion coefficients will be therefore given by

$$
\left|\varphi(t)\right\rangle = \sum_{r_1,r_2,\ldots,r_n=1}^{B_n} \sum_{f_1,f_2,\ldots,f_\nu,\ldots,f_m=0}^{N} \phi\left(r_{1\uparrow},r_{1\downarrow},r_{n\uparrow},r_{n\downarrow},f_{1\uparrow},f_{1\downarrow},\ldots,f_{m\uparrow},f_{m\downarrow}\right) \tag{25}
$$

The generalized JCPM Hamiltonian $\mathbb{H}$ should be also similarly updated:

$$
\mathbb{H} = \hat{\mathbb{H}}_0 + \hat{\mathbb{H}}_{\mathbf{r}\cdot\mathbf{E}} + \hat{\mathbb{H}}_{\mathbf{r}\cdot\mathbf{r}} \tag{26}
$$

$$
\hat{\mathbb{H}}_0 = \sum_{n,i}\left(E^n_{i\uparrow}\hat{\sigma}^n_{i\uparrow} + E^n_{i\downarrow}\hat{\sigma}^n_{i\downarrow}\right) + \sum_{\nu}\left(\hbar\Omega_{\nu\uparrow}\hat{a}_{\nu\uparrow}{}^{\dagger}\hat{a}_{\nu\uparrow} + \hbar\Omega_{\nu\downarrow}\hat{a}_{\nu\downarrow}{}^{\dagger}\hat{a}_{\nu\downarrow}\right) \tag{27}
$$

$$
\begin{aligned}
\hat{\mathbb{H}}_{\mathbf{r}\cdot\mathbf{E}} &= \sum_{n,i<j}\left(\gamma_{nij}\hat{\sigma}^n_{i\uparrow,j\downarrow} + \gamma_{nij}{}^{*}\hat{\sigma}^n_{j\downarrow,i\uparrow}\right)\sum_{\nu}\left(g_{nij\nu\uparrow}\hat{a}_{\nu\uparrow} + g_{nij\nu\uparrow}{}^{*}\hat{a}_{\nu\uparrow}{}^{\dagger}\right) \\
&+ \sum_{n,i<j}\left(\gamma_{nij}\hat{\sigma}^n_{i\downarrow,j\uparrow} + \gamma_{nij}{}^{*}\hat{\sigma}^n_{j\uparrow,i\downarrow}\right)\sum_{\nu}\left(g_{nij\nu\downarrow}\hat{a}_{\nu\downarrow} + g_{nij\nu\downarrow}{}^{*}\hat{a}_{\nu\downarrow}{}^{\dagger}\right)
\end{aligned} \tag{28}
$$

$$\hat{\mathbb{H}}_{\mathbf{r}\cdot\mathbf{r}} = \sum_{n<m} \left( \eta_{nij} \hat{\sigma}^{n}_{i\uparrow,j\uparrow} + \alpha_{nij} \hat{\sigma}^{n}_{i\downarrow,j\downarrow} + \delta_{nij} \hat{\sigma}^{n}_{i\uparrow,j\downarrow} \right) \left( \eta_{mij} \hat{\sigma}^{m}_{i\uparrow,j\uparrow} + \alpha_{mij} \hat{\sigma}^{m}_{i\downarrow,j\downarrow} + \delta_{mij} \hat{\sigma}^{m}_{i\uparrow,j\downarrow} \right) \tag{29}$$

In the first interaction term $\hat{\mathbb{H}}_{\mathbf{r}\cdot\mathbf{E}}$ we note that any atomic transition due to interaction with photon should accompany by a unit change in photon spin. Hence those terms violating the conservation of total spin have been dropped. In the absence of an external magnetic field (or any symmetry breaking perturbation), states having up or down states are double degenerate and the system may be effectively simplified to that of the interaction picture described in Section II. Otherwise, the above sets of equations are minimally needed to fully account for the effect of spin interaction terms.

Now, we proceed in the following to present a full numerical simulation of electron-photon spin entanglement effect.

### A. Full numerical simulation of spin entanglement

As an example of our methodology, the matter photon entanglement observed recently in [64] is simulated here. Following the reported experiment in [64], a quantum optical system consisting of a two-level emitter with two energy states, embedded in a planar photonic micro-cavity, is placed in interaction with a cavity mode occupied with one photon. The emitter is an InAs quantum-dot, that due to the presence of an external magnetic field applied perpendicularly to its growth direction is turned into two coupled Λ-emitter systems. These two emitters consist of separate excited and ground energy states as depicted in Fig. 21. Each of the excited states or trion states have two paired electrons and an unpaired hole ($\left|\uparrow\downarrow\Downarrow\right\rangle$ and $\left|\uparrow\downarrow\Uparrow\right\rangle$) coupled to two ground states which have spin states $\left|\uparrow\right\rangle$ and $\left|\downarrow\right\rangle$ as shown in Fig. 21. The emitters are assumed to be in interaction with a photonic qubit ($H$ and $V$ polarized photons) at the wavelength of 910.10nm. In [64] the energy difference between to emitters due to a magnetic field of 3T corresponds to an angular frequency of $2\pi \times 17.6GHz$. To simulate this system, first by following (24) the ket state of the system is specified as

$$\left|\varphi(t)\right\rangle = \sum_{A_{\uparrow}=g_{\uparrow},g_{\downarrow},e_{\uparrow\downarrow\Uparrow}} \sum_{A\downarrow=g_{\downarrow},g_{\uparrow},e_{\uparrow\downarrow\Downarrow}} \sum_{F_{1\uparrow}=0}^{1} \sum_{F_{1\downarrow}=0}^{1} \phi\left(A,F\right)\left|A\uparrow\right\rangle\left|A\downarrow\right\rangle\left|F_1\uparrow\right\rangle\left|F_1\downarrow\right\rangle \tag{30}$$

Following (25) the describing Hamiltonian for this system is given as

$$\hat{\mathbb{H}}_0 = \sum_{1,i=g\uparrow,g\downarrow,e\uparrow\downarrow\Uparrow,e\uparrow\downarrow\Downarrow} \left(E^1_{i\uparrow}\hat{\sigma}^1_{i\uparrow} + E^1_{i\downarrow}\hat{\sigma}^1_{i\downarrow}\right) + \sum_{\nu}\left(\hbar\Omega_{\nu\uparrow}\hat{a}_{\nu\uparrow}{}^{\dagger}\hat{a}_{\nu\uparrow} + \hbar\Omega_{\nu\downarrow}\hat{a}_{\nu\downarrow}{}^{\dagger}\hat{a}_{\nu\downarrow}\right) \tag{31}$$

$$\begin{aligned}\hat{\mathbb{H}}_{\mathbf{r}\cdot\mathbf{E}} = &\sum_{1,i=g_{\uparrow},g_{\downarrow}<j=e_{\uparrow\downarrow\Uparrow},e_{\uparrow\downarrow\Downarrow}} \left(\gamma_{1ij}\hat{\sigma}^1_{i\uparrow,j\downarrow} + \gamma_{1ij}{}^{*}\hat{\sigma}^1_{j\downarrow,i\uparrow}\right)\left(g_{1ij\nu_1\uparrow}\hat{a}_{\nu_1\uparrow} + g_{1ij\nu_1\uparrow}{}^{*}\hat{a}_{\nu_1\uparrow}{}^{\dagger}\right)\\ &+\sum_{1,i<j}\left(\gamma_{1ij}\hat{\sigma}^1_{i\downarrow,j\uparrow} + \gamma_{1ij}{}^{*}\hat{\sigma}^1_{j\uparrow,i\downarrow}\right)\left(g_{1ij\nu_1\downarrow}\hat{a}_{\nu_1\downarrow} + g_{1ij\nu_1\downarrow}{}^{*}\hat{a}_{\nu_1\downarrow}{}^{\dagger}\right)\end{aligned} \tag{32}$$

in which $i, j$ denote the energy states, in such a way that in any term $i$ refers to an energy level lower than that of $j$. $\nu_1\uparrow$ and $\nu_1\downarrow$ refer to the cavity modes occupied by $H$ and $V$ polarized photons, respectively. Finally, $\hat{\mathbb{H}}_{\mathbf{r}\cdot\mathbf{r}}$ has to be set to zero because it describes the interaction between two physically separate emitters (e.g. in two quantum dots), and there is no such interaction term.

Now, by using the Hamiltonian defined in (31) and (32), and following the method of solving Schrödinger's equation (1) similar to what has been done in the previous section, the coefficient matrix and time-dependent ket of the system can be computed without any approximation.

In this simulation, initially the system is supposed to be in the ground state, with the spin down and up interacting with one cavity mode occupied by the $H$ polarized photon in weak coupling regime. The presence probability of the system being in this ket state is measured and plotted in Fig 21 according to (12). In order to study the entanglement, the concurrency parameter is also computed as explained extensively elsewhere [56,63].

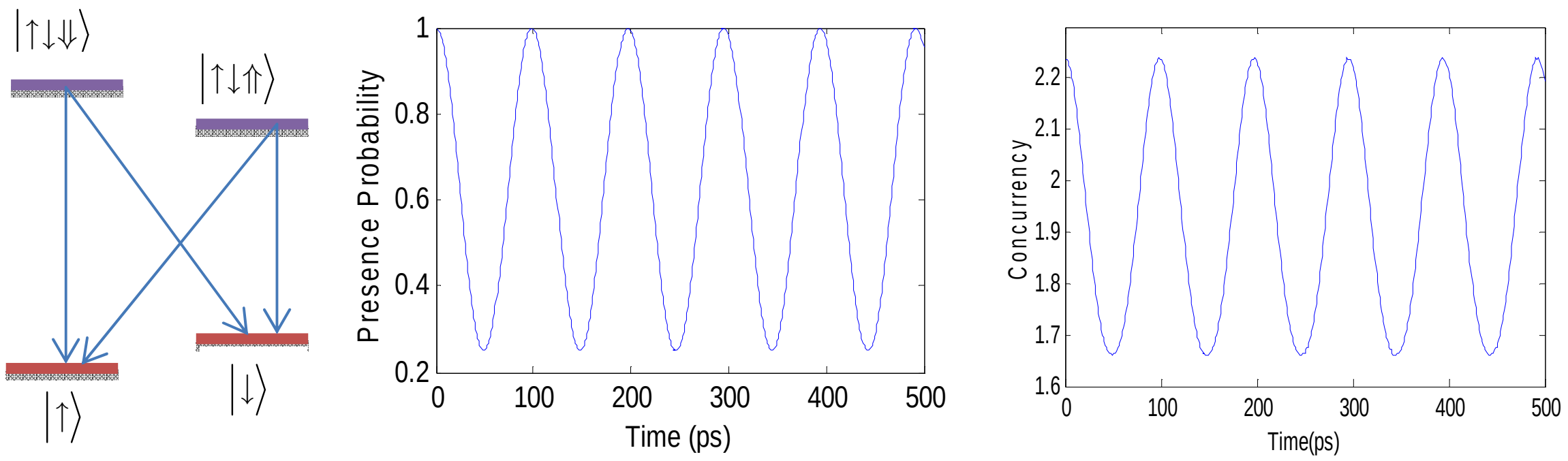

FIG. 21. Two coupled Λ-emitter systems. Presence probability and concurrency versus time are calculated respectively.

As it was expected from experimental observations [64], we may observe a sinusoidal behavior for the probability with a time-scale of the order of 100 pico-second. Based on the simulation data, the interaction regime falls in the weak coupling. Similarly, the variations of concurrency also attain the same time-scale of oscillations. Nearly sinusoidal oscillations of this parameter confirms the periodic build-up and destruction of entanglement between quantum-dot's electron spin and photon's polarization.

## V. CONCLUSION

In this paper, the general behavior of CQED of complex systems under different coupling regimes was analyzed. Mathematically we tackled the most general quantum optical system consists of an arbitrary number of light emitters interacting with an arbitrary number of cavity modes. We presented how to specify the general time dependent state of the system, how to provide initial conditions and to solve the system without any approximation in time-domain in Schrödinger picture. Next, we presented expressions for measuring presence probabilities, expectation value of field operators, atomic operators, and commutators. We have developed an extensive MATLAB code to produce the necessary initial conditions and solve the system. We also presented and discussed two example systems in details. We confirmed that RWA may not be used in ultrastrong coupling. We furthermore have observed, for the first time, a chaotic behavior in ultrastrong coupling regime accompanied by multi-step and random-like abrupt phase changes.

## Acknowledgments

This work was supported in part by Iranian National Science Foundation (INSF) under Grant 89001329.